\documentstyle[12pt]{article}
\if@twoside
 \else
 \fi

 \parskip \medskipamount
\newcommand{\be}{\begin{eqnarray}}

\newcommand{\ee}{\end{eqnarray}}
\begin{document}
%\twocolumn

%\preprint{??}
\begin{center}
\vspace{24pt}
{\large \bf Anomalous chirality fluctuations  in 
the initial stage of heavy ion collisions and parity odd bubbles}

\vspace{36pt}

{\sl D. Kharzeev}$^{a}$, {\sl A. Krasnitz}$^{b}$ and
{\sl R. Venugopalan}$^{a,c}$

\vspace{24pt}\noindent
$^a$ Physics Department, Brookhaven National Laboratory, Upton, NY 11973, 
USA\\
{E-mail: kharzeev@bnl.gov} \\

\vspace{12pt}\noindent
$^b$ CENTRA and Faculdade de Ci\^encias e Tecnologia,\\ Universidade do 
Algarve,
Campus de Gambelas, P--8000, Faro, Portugal \\
{E-mail: krasnitz@ualg.pt}

\vspace{12pt}\noindent
$^c$ RIKEN-BNL Research Center, Brookhaven National Laboratory, \
Upton, NY 11973, USA\
{E-mail: raju@bnl.gov}

\end{center}
\vspace{12pt}

\vfill

\begin{center}
{\bf Abstract}
\end{center}

We compute numerically the topological charge distribution in the
initial stage of a high energy heavy ion collision. This charge
distribution is generated by Chern-Simons number fluctuations
associated with the dynamics of strong classical fields in the initial
state. The distribution is found to be quite narrow at RHIC and LHC
energies reflecting a small value of the topological susceptibility. 
Thus the effective potential of classical fields is 
shallow in the $\theta$-direction likely creating favorable
conditions for the subsequent generation of P-odd bubbles.

\vspace{12pt}

\noindent

\bigskip



\vfill

\newpage

%\vspace*{0.2cm} 

\section{Introduction}

In high energy heavy ion collisions, QCD matter is produced at very
high energy densities. An issue that has aroused considerable
interest is the possibility ~\cite{KPT,KP} of creating metastable
states that break the discrete global symmetries of parity ($P$) and
charge-parity ($CP$) satisfied by the QCD
Lagrangian~\footnote{This is true in the absence of a $\theta$--term
which would break $P$ and $CP$ explicitly.  We will assume that
$\theta=0$ here.}. There is a theorem by Vafa and Witten which states
that there can be no stable parity violating phase of
QCD~\cite{VafaWitten}. The conditions of applicability of this theorem
have attracted much attention recently~\cite{AG,SS,Ji}. 
The Vafa-Witten theorem does not however preclude the
possibility of forming metastable domains that violate parity (and
charge--parity)~\cite{KPT}. Furthermore, it has been suggested
recently that the Vafa--Witten theorem may not be applicable at finite
temperature~\cite{TomCohen}.

Several experimental signatures have recently been suggested that
would be sensitive to the formation of these $P$ and $CP$ odd
metastable states~\cite{KP}. Experimental searches are currently
underway at RHIC. A preliminary report on their status was discussed
at Quark Matter 2001~\cite{QM2001}.

In the discussions of Refs.~\cite{KPT,KP}, the formation of $P$ and $CP$--odd
domains is studied in the context of the restoration of the axial
$U(1)$ symmetry at finite temperature.  A non--linear $\sigma$--model
which incorporates the breaking of the $U(1)$ axial
symmetry~\footnote{For a recent study of $U_A$(1) restoration in the
linear sigma model, see Ref.~\cite{Lenaghan}.}, the Di
Vecchia-Veneziano-Witten Lagrangian~\cite{deVVW}, (VVW) is constructed
and it is shown that there exist metastable states (corresponding to
local minima of the potential) which spontaneously break $P$ and
$CP$~\cite{KPT}. The likelihood of forming long-lived metastable states
depends on the coefficient $a\sim m_{\eta^\prime}^2$ of the anomaly
term in the VVW Lagrangian -- they are more likely when $a(T)\rightarrow
0$ as the temperature approaches the deconfinement temperature, 
$T\rightarrow T_c$. There is evidence from lattice simulations that this is
the case. The anomaly term is proportional to
the topological susceptibility and lattice data suggest that the
topological susceptibility for both quenched and full QCD drops
rapidly in the vicinity of $T_c$ and is consistent with zero by
$T>1.5 T_c$ ($1.2 T_c$) for $N_f=2$
($N_f=4$)~\cite{DiGiacomo}. Further, large $N_c$ analyses suggest
that $a(T)\rightarrow 0$ as $T\rightarrow T_c$~\cite{KPT,KP,PT,Affleck}.

In Ref.~\cite{KP}, the decay of $P$ and $CP$ odd
metastable states into pions is described by the Wess--Zumino--Witten
term~\cite{WZW} in the chiral Lagrangian.  This term preserves the net
chirality or handedness generated in the $P$ and $CP$ odd domains. 
The magnitude of the net asymmetry produced by this mechanism 
in heavy ion collisions is estimated to be $\sim 10^{-3}$~\cite{KP}.
The signatures of P and CP odd domains were investigated also in Refs.
\cite{BFZ,ABD,Chaud}, and include an enhancement in the $\eta$ and $\eta'$ yields,
generically associated with the $U_A(1)$ restoration \cite{KKM}.

It is of great interest to investigate how the topological
susceptibility is affected by the dynamical conditions generated in
the early stage of a heavy ion collision. 
Immediately after a high energy
nuclear collision, as a consequence of the phenomenon of gluon saturation~\cite{GLR}, 
large color electric and magnetic fields are
generated. These can in principle produce a large amount of
topological charge through the mechanism of sphaleron transitions
already in the initial stage of the heavy ion collision.
However, as we will argue below, the boost invariance of the initial
stage of the heavy ion collision suppresses sphaleron transitions. The
primary mechanism for the generation of topological charge at the
early stage is by fluctuations of the color electric and magnetic
fields. For a translationally invariant system, the fluctuations in
the topological charge determine the topological susceptibility.

In this paper, we estimate the size of topological charge fluctuations
in the early stage of a heavy ion collision.  Our computation is
performed within the framework of a classical effective field theory
(EFT)~\cite{MV,JKLW} approach to small $x$ physics. In this approach,
parton distributions in a nucleus {\it before the collision} form a
color glass condensate (CGC)~\cite{RajGavai,ILM}. The shattering of 
the color glass condensate and subsequent multiparticle production 
in high energy scattering has been discussed 
for eA~\cite{MV}, pA~\cite{Adrian} and for nuclear collisions~\cite{KMW,KharzeevNardi}.
For nuclear collisions, the problem can be
formulated in the EFT~\cite{KMW} and the temporal evolution of gluons
produced after the collisions can be solved for
numerically~\cite{AR99}. The initial energy~\cite{AR00} and
number~\cite{AR01} of produced gluons have been computed previously in
this approach. The CGC idea has been shown to be consistent with many
features of data from heavy ion collisions~\cite{KharzeevNardi,JuergenLarry}.

We find in the CGC model that the fluctuations in topological charge 
are quite small at RHIC and LHC energies. The root mean squared topological charge 
is only one unit per two units of rapidity at RHIC energies and one unit per 
unit of rapidity at LHC energies. Our results suggest that the form of the effective 
potential in the $\theta$--direction is quite shallow at early times. This result makes 
favorable the formation of $P$ and $CP$ odd bubbles at later times -- namely, the 
scenario that was first discussed in Refs.~\cite{KPT,KP}.

Recently instanton/sphaleron approaches to multiparticle production
have been investigated~\cite{KKL,SZ,KKL1,Shuryak}. In Ref.~\cite{KKL1}
it was shown that in the initial stage of nuclear collisions instanton
transitions (and therefore the topological susceptibility) are
severely suppressed by strong classical gluon fields.  Shuryak has
suggested that the dominant mechanism for particle production in
nuclear collisions may be through the decay of
sphalerons~\cite{Shuryak}.  If so, a likely consequence is that the
ensuing sphaleron transitions will generate large $P$ and $CP$
violation already in the initial stage of the nuclear collision. In our approach, 
however, sphaleron transitions are suppressed by
strict boost invariance. Sphaleron transitions, if they occur, will
take place at a later stage when the strict boost invariance of the
initial state is lost. We have estimated the topological charge generated by 
sphaleron transitions in a hot gluon plasma and find, for RHIC energies, that it 
is large compared to the topological charge generated in the initial stage by 
classical fields.

The paper is organized as follows. In section 2, we briefly review the 
CGC picture of classical gluon production in heavy ion collisions. 
The  no--go theorem for sphaleron transitions for a strictly boost invariant 
system is derived
in section 3. Numerical results for the Chern--Simons charge are 
presented in section 4. Its relation to empirical observables is 
also discussed there. In section 5, we discuss estimates for 
$P$ and $CP$ violation from sphaleron transitions at finite temperature in 
equilibrium hot gluodynamics. We also comment on the implications 
of these results (versus those in the boost invariant CGC picture) 
for the heavy ion experiments.

\section{Classical gluon production in nuclear collisions}
\vskip 0.2in

In a high energy heavy ion collision, the dynamics of the central rapidity 
region is determined by the small $x$ modes in the individual 
nuclei before the collision. At small $x$, the gluon density grows 
rapidly -- however, repulsive effects soon become important and slow down the 
growth of the gluon density. This phenomenon is called saturation and 
has been discussed extensively in the literature~\cite{GLR}. The occupation 
number of gluons in the saturated regime is proportional to 
$1/\alpha_s(\Lambda_s)$ evaluated at a saturation scale $\Lambda_s$. If 
$\Lambda_s\gg \Lambda_{QCD}$, the occupation number is large and classical 
methods may be applied to study parton distributions in the 
nuclei~\cite{MV,JKLW}. It has been shown recently that a
RG-improved generalization of this classical effective field theory (EFT) 
reproduces several key results in small-$x$ QCD~\cite{JKLW,ILM}.
The classical EFT was first applied to the study of collisions of large
nuclei by Kovner, McLerran and Weigert~\cite{KMW}. The model, as applied
to nuclear
collisions, may be summarized as follows. The colliding nuclei are idealize
to travel along the light cone
The high-$x$ and the low-$x$ modes in the nuclei are treated separately. Th
former corresponds to valence quarks and hard sea partons and
are considered recoilless sources of color charge
For each of the large Lorentz-contracted nuclei (for simplicity, we will
consider only collisions of identical nuclei), this results in a static
Gaussian distribution of their color charge density $\rho_{1,2}$ in th
transverse plane
$$P\left([\rho]\right)\propto\, {\rm exp}\left[-{1\over{2\Lambda_s^2}}
\int{\rm d}^2r_t\rho_{1,2}^2(r_t)\right].$$
The variance $\Lambda_s$ of the
color charge distribution is the only dimensionful parameter of the
model, apart from the linear size $L$ of the nucleus. For central
impact parameters, $\Lambda_s$ can be estimated in terms of single-nucleon 
structure
functions~\cite{GyulassyMcLerran}. It is assumed, in addition, that the
nucleus is infinitely thin in the longitudinal direction. Under this
simplifying assumption, the resulting gauge fields are explicitly 
boost-invariant.

The small $x$ fields are then described by the classical Yang-Mills equations
\begin{equation}
D_\mu F_{\mu\nu}=J_\nu\label{eqmo}
\label{YM}
\end{equation}
with the random sources on the two light
cones:
$J_\nu=\sum_{1,2}\delta_{\nu,\pm}\delta(x_\mp)\rho_{1,2}(r_t).$
The two signs correspond to two possible directions of motion along the
beam axis $z$. As shown by Kovner, McLerran and Weigert (KMW)~\cite{KMW}, 
low-$x$ fields
in the central region of the collision obey sourceless Yang-Mills equations
(this region is in the forward light cone of both nuclei) with the initial
conditions in the $A_\tau=0$ gauge given by
\begin{equation}
A^i=A^i_1+A^i_2;\ \ \ \ A^\pm=\pm{{ig}\over 2}x^\pm[A^i_1,A^i_2].
\label{incond}\end{equation}
Here the pure gauge fields $A^i_{1,2}$ are solutions of (\ref{eqmo}) for each
of the two nuclei in the absence of the other nucleus.

In order to obtain the resulting gluon field configuration at late
proper times, one needs to solve (\ref{eqmo}) with the initial condition 
(\ref{incond}).
Since the latter depends on the random color source,
averages over realizations of the source must be performed. 
KMW showed that in perturbation theory
the gluon number distribution by transverse momentum (per unit rapidity)
suffers from an	infrared divergence and argued that the distribution must
have the form
\begin{equation}
n_{k_\perp}\propto{1\over\alpha_s}\left({{\Lambda_s}\over k_\perp}\right)^4\ln\left({k_\perp\over{\Lambda_s}}\right) \label{dpt}\end{equation}
for $k_\perp\gg\Lambda_s$. 
The log term clearly indicates that the perturbative description breaks down
for $k_\perp\sim\Lambda_s$. 

A reliable way to go beyond perturbation theory is to re-formulate the
EFT on a lattice by discretizing the transverse plane. The resulting
lattice theory can then be solved numerically. We shall not dwell here
on the details of the lattice formulation, which is described in
detail in Ref.~\cite{AR99,AR00}.  Keeping in mind that $\Lambda_s$ and
the linear size $L$ of the nucleus are the only physically interesting
dimensional parameters of the model~\cite{RajGavai}, we can write any
dimensional quantity $q$ as $\Lambda_s^df_q(\Lambda_s L)$, where $d$
is the dimension of $q$. All the non-trivial physical information is
contained in the dimensionless function $f_q(\Lambda_s L)$.  We can
estimate the values of the product $\Lambda_s L$ which correspond to
key collider experiments. Assuming Au-Au collisions, we take $L=11.6$
fm (for a square nucleus!) and estimate the saturation scale
$\Lambda_s$ to be 2 GeV for RHIC and 4 GeV for
LHC~\cite{GyulassyMcLerran}.  Also, we have approximately $g=2$ for
energies of interest. The rough estimate is then $\Lambda_s L\approx
120$-$150$ for RHIC and $\Lambda_s L\approx 240$-$300$ for LHC.  Since
the gluon distribution in nuclei is not known to great precision,
there is a considerable systematic uncertainty in these estimates.

This model has been applied recently to study classical gluon production.
The energy and number
distributions have been computed numerically, and the dependence of these
quantities on $\Lambda_s$ has been determined~\cite{AR00,AR01}. We 
will now apply the model to compute the Chern Simons number generated in
the early stages of a heavy ion collision.

\section{Boost invariance and Chern Simons number}
\vskip 0.2in

In general, the Chern-Simons (CS) number changes under a gauge
transformation by an integer equal to the winding number of that
transformation. Since the potential of the theory is invariant under
all gauge transformations, there exists a direction in the functional
space along which the potential is periodic. The system can move along
that direction arbitrarily far from its initial configuration. Such motion
generates 
arbitrarily large changes of the CS number or, equivalently, 
arbitrarily large values of topological charge.  Dynamical, real time, 
processes corresponding to nontrivial gauge transformations are called
sphaleron transitions. In a large space-time volume these transitions
proceed independently in causally unrelated regions. If, in addition,
the space-time is translation-invariant, the topological charge for a
large space-time volume is a random walk: 
\be
\langle\left(N_{\rm cs}(t)-N_{\rm cs}(t')\right)^2\rangle=\Gamma V(t-t')\,,
\label{TS}
\ee
where $\Gamma$ is known as the sphaleron transition rate, or as 
the topological susceptibility. Here, and in the following, $\langle ... \rangle$
means an ensemble average: the thermal ensemble in thermal equilibrium,
and the ensemble of (central) nuclear collision events in the case of interest
here.
This random walk behavior was indeed observed in numerical simulations of SU(2) 
\cite{AmbjornKrasnitz} and SU(3) \cite{Moore} Yang-Mills theories at high
temperature.

The dynamics of the CS number in the central region of a nuclear
collision differs from the generic case in two ways. Firstly, the time
translation invariance is broken by the instance of the collision itself. 
Secondly, and more importantly, there are no nontrivial boost-invariant
gauge transformations. Consequently, for boost-invariant
configurations, such as those in the central region, the potential is not 
periodic, and there can be no sphaleron transitions. 
Instead, one expects the CS number to fluctuate in the vicinity of
zero. 

In order to make this second point explicit, we shall now examine the form of 
the CS number functional for boost-invariant configurations, 
and show it to be invariant under {\it all} boost-invariant gauge 
transformations. It will then follow that, in this special case, the CS number
cannot be changed by a dynamical process equivalent to a gauge transformation,
{\it i.e.}, that sphaleron 
transitions are prohibited for a boost invariant system.

It is convenient for us to use the coordinates $(\vec r_t,\tau,\eta)$. Here
$\tau = \sqrt{2 x^+ x^-}$ is the proper time, $\eta = \ln(x^+/x^-)/2$ 
is the space-time rapidity, and $\vec r_t\equiv(x,y)$ is the position in the 
transverse plane. For the components of the gauge potential, the 
boost invariance assumption (see the discussion
immediately prior to Eq.~\ref{YM}) means that
\be
\vec A_t(\tau,x_t,\eta)=\vec A_t(\tau,x_t)\,\,;\,\, A_{\eta}(\tau,x_t,\eta)
\equiv\Phi(\tau,x_t)\, ,
\label{GF}
\ee
and the system effectively becomes (2+1)-dimensional, with $A_{\eta}$ becoming
an adjoint scalar field.
It is also convenient to remove the fourth component of $A$ by imposing the 
gauge condition
$A^\tau=0$.

Now consider the quantity
\be
\nu = \frac{1}{16\pi^2}\,\int {\rm d}^2 \, x_t \Phi^a B^a \, ,
\label{CSN}
\ee
where 
\be
B^a = F^a_{xy}=\partial_xA^a_y-\partial_yA^a_x+g\epsilon^{abc}A^b_xA^c_y
\label{ba}\ee
is the component of the magnetic field
perpendicular to the transverse plane. This quantity is obviously
invariant under all rapidity independent gauge transformations.
In order to verify that $\nu$
is the CS number per unit rapidity, we will compute its proper time
derivative (denoted by a dot in the following)
and show it to be equal to the topological charge density integrated
over the transverse plane. We recall that the topological charge density
is given by
\be
\frac{1}{16\pi^2}\vec E\cdot\vec B\, ,\label{tcd}\ee
$\vec E^a$
and $\vec B^a$ being the color electric and magnetic fields, respectively;
the anomaly equation relates this quantity to the divergence of the chiral
current.

To this end, it suffices to write out $\dot\nu$ explicitly, using (\ref{ba}),
and integrate the terms containing $\partial_x\dot A^a_y$ and
$\partial_y\dot A^a_x$ by parts in the transverse plane, assuming
that the corresponding boundary terms vanish. The result is
\be
\dot\nu={1\over{16\pi^2}}\int{\rm d}^2x_t\left(\dot\Phi^aB^a
+\dot A^a_xD^a_y\Phi-\dot A^a_yD^a_x\Phi\right),
\label{dotnu}\ee
where $D^a_i\Phi$ is the covariant derivative: 
$D^a_i\Phi\equiv\partial_i\Phi^a+\epsilon^{abc}A^b_i\Phi^c$. To complete the
proof, we note that ({\it a}) for boost invariant configurations, the
integrand of (\ref{dotnu}) is equal to $\tau\vec E^a\cdot\vec B^a$;
and ({\it b}) the space-time volume element in the $\tau,\eta,\vec r_t$ 
coordinates is ${\rm d}^4x=\tau{\rm d}\eta{\rm d}\tau{\rm d}^2x_t$.
Hence we have shown that the proper time variation of $\nu$ in a boost invariant
process gives the topological charge per unit rapidity of the process. If any
such process is equivalent to a gauge transformation, it leaves $\nu$ unchanged,
the corresponding topological charge vanishes, and no sphaleron transitions are 
possible. 

In the EFT description of nuclear collisions the scalar field $\Phi$ vanishes 
at $\tau=0$ \cite{AR99}, and, therefore, $\nu(\tau=0)=0$. In other words, the
topological charge of the process at any given $\tau$ is simply $\eta\nu(\tau)$,
where $\eta$ is the rapidity extent of the process. Since the EFT has a
parity-even action, both signs of $\nu(\tau)$ are equally probable, and the 
ensemble average of $\nu(\tau)$ vanishes. Our principal object of interest is 
then $\langle\nu^2(\tau)\rangle$. In the absence of sphaleron transitions, we
cannot expect the proper time evolution of $\nu$ to resemble a random walk.
Indeed, as we shall see in the next section, $\langle\nu^2(\tau)\rangle$ 
approaches a constant at large $\tau$.

 Strictly speaking, because of the strict 
boost invariance built into our approach,  
$\langle\nu^2(\tau)\rangle$ per unit space-time volume cannot be interpreted
as the topological susceptibility.
Nevertheless, the computed $\langle\nu^2(\tau)\rangle$ is a measure  
of the contribution of classical fields to the topological susceptibility.  
As will be discussed further in Section 5, the likelihood of forming 
long-lived P and CP odd domains depends sensitively on this quantity.

\begin{figure}
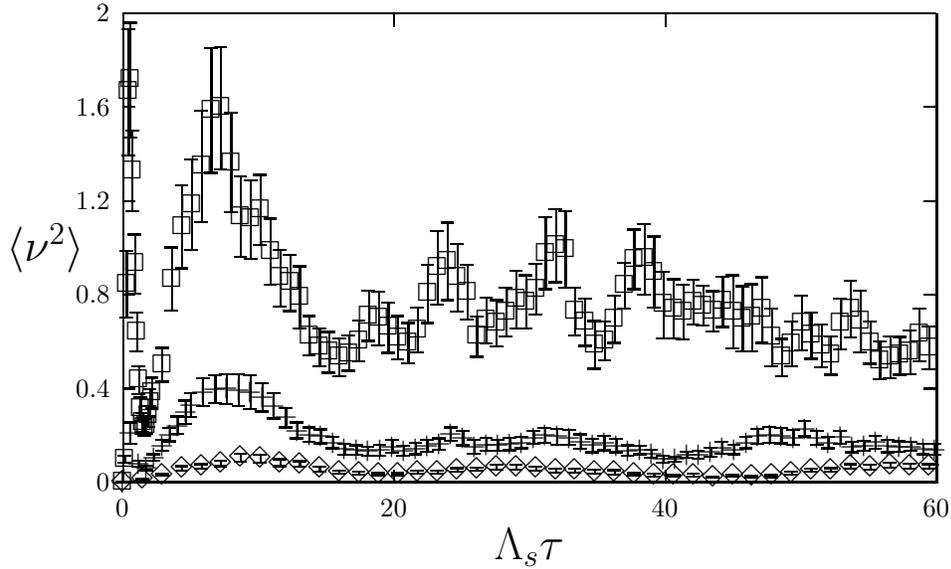

% GNUPLOT: LaTeX picture
\setlength{\unitlength}{0.240900pt}
\ifx\plotpoint\undefined\newsavebox{\plotpoint}\fi
\sbox{\plotpoint}{\rule[-0.200pt]{0.400pt}{0.400pt}}%

\caption{Average squared topological charge {\it vs} the proper time (in units
of $\Lambda_s^{-1}$) for $\Lambda_sL=74.2$ (diamonds), 148.4 (pluses), and
297 (squares). The latter two values of $\Lambda_s L$ correspond to the
RHIC ($\Lambda_s=2$ GeV) 
and LHC ($\Lambda_s=4$ GeV) regimes respectively.}
\end{figure}

\begin{figure}
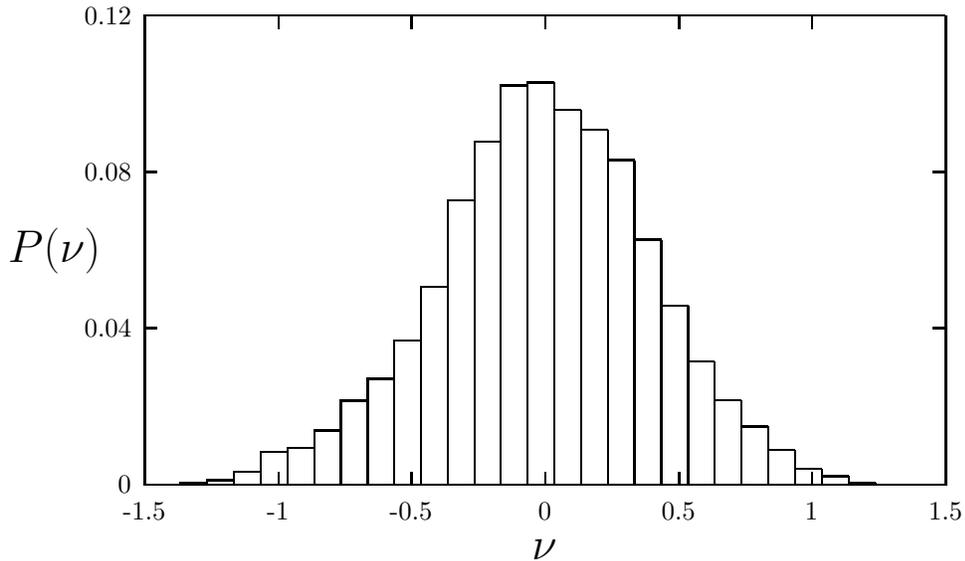

% GNUPLOT: LaTeX picture
\setlength{\unitlength}{0.240900pt}
\ifx\plotpoint\undefined\newsavebox{\plotpoint}\fi
\sbox{\plotpoint}{\rule[-0.200pt]{0.400pt}{0.400pt}}%
%
\caption{Probability distribution of $\nu$ in the proper time interval
$30\leq\Lambda_s\tau\leq 65$ for $\Lambda_s L=148.4$. These corresponds 
to $3$ fm $\leq \tau \leq 6$ fm for an estimated value $\Lambda_s=2$ GeV for
RHIC energies.}
\end{figure}

\section{Numerical results}

We use a lattice formulation of the EFT developed in \cite{AR99,AR00,AR01}.
As explained in these references, the key dimensionless parameter for those
observables which exist in the continuum limit is $\Lambda_s L$, where $L$ is 
the radius of the nucleus (assuming a square nucleus, $L^2 = \pi R^2$).
The value of $\Lambda_s$, expressed in units of the lattice spacing $a$, 
indicates how close the system is to the continuum limit.
In the current study, for an SU(2) gauge theory~\footnote{We study an
SU(2) gauge theory since it is simpler to simulate numerically. The
physically more interesting case of an SU(3) gauge theory was studied
recently and gluon number distributions computed~\cite{KNV}. No significant
qualitative differences from the SU(2) case were found for the energies of
interest.}
we set $\Lambda_s a=0.29$. The CS number $\nu$ is a dimensionless
quantity, whose ultraviolet properties are similar to those of the total gluon
number $N$, studied in \cite{AR01}. In that study, we determined that 
discretization artifacts are very small for $N$ at $\Lambda_s a=0.29$.

We examined the proper time evolution of $\nu$ for three values of
$\Lambda_sL$, namely, 74.2, 148.4, and 297. The last two of these
correspond roughly to central Au-Au collisions in the RHIC and LHC
regimes, respectively. In each of the three cases, we generated about
80 independent initial configurations and solved the classical
equations of motion for the duration of proper time of at least
$60/\Lambda_s$ (this corresponds to $6$ fm for RHIC).  The average
proper time histories of $\nu^2$ in each case are presented in Figure
1. Evidently, $\nu$ varies over a much longer time scale, compared to
other global quantities, such as the total energy and the total
particle number \cite{AR00,AR01}. Nevertheless,
$\langle\nu^2(\tau)\rangle$ does approach a constant value at late
$\tau$. That constant value is a rapidly growing function of 
$\Lambda_s L$. Since the dynamics considered here is exactly
boost-invariant, $(\langle\nu^2(\tau)\rangle)^{1/2}$ is a measure of
the anomalous chirality violation per unit rapidity in the initial stage of a 
central collision. 

To estimate the magnitude of the effect, we took the proper time average 
of $\langle\nu^2(\tau)\rangle$ between $\Lambda_s\tau=30$ and 
$\Lambda_s\tau=65$. The results are summarized in Table 1. If we take the
experimental range of two units of rapidity, it follows that for RHIC 
we have approximately one unit of the topological charge squared 
for two units of rapidity while at LHC one may have two units of topological 
charge squared in the same rapidity interval. Note that $\langle\nu^2\rangle$
is approximately proportional to $(\Lambda_s R)^2$, {\it i.e.}, to the 
transverse area in units of the nonperturbative scale $\Lambda_s$. This means
that dynamical processes contributing to the topological charge become
uncorrelated at spatial separations exceeding $\Lambda_s$, and the resulting
fluctuations of $\nu$ add up randomly.
\begin{table}[h]
\centerline{\begin{tabular}{|lrrr|} \hline
$\Lambda_s R$ & 41.9 & 83.7 & 167.5 \\
$\langle\nu^2\rangle$ & $0.046\pm 0.004$ & $0.16\pm 0.01$ & $0.69\pm 0.05$ \\
\hline
\end{tabular}}
\caption{The average value of $\langle\nu^2\rangle$, tabulated as a function
of the dimensionless parameter $\Lambda_s R$.}
\end{table}

As we have explained in the preceding sections, there is no CS
number diffusion in a boost invariant setting, nor is there a ground state
degeneracy. It is for this reason that $\langle\nu^2(\tau)\rangle$ 
approaches a constant at large $\tau$, instead of growing linearly, as would
be the case if sphaleron transitions were permitted. Also, as Figure 2
illustrates, there is no special significance to integer values of $\nu$ other
than zero, which in a general case correspond to degenerate minima of the
potential.

\vskip 0.2in

\section{Comparison with other estimates of the topological 
susceptibility and experimental ramifications}
\vskip 0.2in

Thus far, we have assumed that strict boost invariance holds in heavy ion 
collisions at very high energies. If strict boost invariance does not hold, 
sphaleron transitions are allowed. Whether sphaleron transitions turn on 
slowly or rapidly when strict boost invariance is relaxed is not clear. A way 
to answer this question is to introduce a small 
perturbation $a(x_t,\eta,\tau)$, namely,
\be
\tilde{A}(x_t,\eta,\tau) = A(x_t,\tau) + a(x_t,\eta,\tau) \, ,
\ee
in the Yang--Mills equations and to study how it evolves. While the Yang-Mills 
equations equations can be linearized with respect to $a(x_t,\eta,\tau)$,
such a study has to be performed numerically, since $A(x_t,\tau)$ is not
known analytically~\cite{KNV1}.

Let us assume that the system does thermalize and strict boost invariance
of field configurations 
is lost~\footnote{Boost invariance may still be preserved when one averages 
over events.}. We then need an estimate of the sphaleron transition rate in a
hot gluon plasma. To simplify the setup, we shall first ignore dynamical
fermions and assume that the temperature $T$ is much larger than any other
dimensional scale in the problem. It then follows from dimensional
considerations~\cite{ArnoldMcLerran} that the rate is proportional to $T^4$
However, the dependence of the rate on the coupling constant $g$ is less
obvious. 
In the weak-coupling regime the sphaleron transitions are dominated by soft
modes, with momenta of the order of $g^2T$. The time scale for the transition
depends on how important is the interaction between these degrees of freedom
and hard thermal modes. A series of perturbative studies~\cite{ASY,Dietrich},
performed at extremely weak coupling (${\rm ln}(1/g)\gg 1$), suggest that 
this interaction sets the time scale for sphaleron transitions to $(g^4T)^{-1}$
up to logarithmic corrections. Then the coupling constant dependence is
$\Gamma\propto\alpha_s^5T^4$, and a recent numerical estimate gives~\cite{Moore}
\be
\Gamma_{sph} = 108\, \alpha_s^5 \,T^4 \, 
\label{rate2}\ee
On the other hand, there are indications that, away from
the asymptotically weak coupling regime, the relevant time scale for
the soft modes is $(g^2T)^{-1}$~\cite{AAK}. Should this be the case, the
numerical estimate~\cite{Moore} should be re-interpreted to give
\be
\Gamma_{sph} \approx 8\alpha_s^4 \,T^4 \, .
\label{rate1}
\ee
In the following, we will use both estimates. 

For a hot plasma (without dynamical fermions) 
the potential is periodic and one generates topological charge via a 
random walk. One has
\be
\langle \Delta Q_5^2\rangle = \int_0^\tau d^4 x\, \Gamma_{sph} \, ,
\label{tpch}
\ee
where $d^4 x= d^2 x_t \tau d\tau d\eta$.
In an expanding boost invariant plasma, 
the temperature decreases with time. For an isentropic 
expansion~\cite{Bjorken} one has $T\propto \tau^{-1/3}$. More precisely, 
\be
T = \left[{90\over 4\pi^2}\,{dS\over d\eta}\,{1\over {L^2 g_d}}\, 
{1\over \tau}\right]^{1/3} \, ,
\label{temp}
\ee
where $g_d$ is the degeneracy factor and $dS/d\eta$ is the entropy per unit 
rapidity.
From Eq.~\ref{tpch} and Eq.~\ref{temp}, we obtain
\be
{d\langle \Delta Q_5^2\rangle\over d\eta} = 1.5 A\,\alpha_s^n\,
\left({\tau_f\over L}\right)^{2/3}\, \left[{dS\over d\eta}\,{1\over g_d}\,
{90\over 4\pi^2}\right]^{4/3} \, ,
\label{tpch1}
\ee
where $A$ and $n$ are correspond to the different coefficient and power 
respectively in the two sphaleron transition rates.
At RHIC energies, $dS/d\eta\sim 3.6\times 1000$, $\alpha_s\sim 0.3$, 
$\tau_f\approx L\sim 10$ fm, $g_d=16$. Then from the sphaleron transition 
rate in Eq.~\ref{rate2}, and from Eq.~\ref{tpch1}, we find for an expanding 
plasma,
\be
{d\langle \Delta Q_5^2\rangle^{(1)}\over d\eta}|_{RHIC} \approx 1600\,\,;\,\,
{d\langle \Delta Q_5 \rangle_{RMS}^{(1)}\over d\eta}|_{RHIC}\approx 40 \, .
\ee
Here RMS denotes root mean square.
Similarly, for the sphaleron transition rate in Eq.~\ref{rate1}, and 
Eq.~\ref{tpch1}, we have
\be
{d\langle \Delta Q_5^2\rangle^{(2)}\over d\eta}|_{RHIC} \approx 400\,\,;\,\,
{d\langle \Delta Q_5 \rangle_{RMS}^{(2)}\over d\eta}|_{RHIC}\approx 20 \, .
\ee
Though these estimates of the topological charge generated by sphaleron 
transitions differ from each other
they both are significantly larger than the 
topological charge squared obtained from boost invariant classical fields at 
RHIC (and LHC). In this paper, we have considered the case of pure gluodynamics; 
the effects due to dynamical fermions had been addressed in Refs.~\cite{MMS,roberge,kkp}. 

We now come to the phenomenological implications of our results. As
discussed previously, the likelihood of long-lived P and CP violating
metastable states in heavy ion collisions~\cite{KPT,KP} depends
sensitively on the coefficient $a$ of the anomaly term in the VVW
Lagrangian. This coefficient is proportional to the topological
susceptibility. The likelihood of metastable states is greater for
smaller values of $a$. Lattice and large $N_c$ calculations suggest
that the topological susceptibility is indeed small for $T>T_c$
The previous estimates did not include
the possibility of contributions from fluctuations in classical fields
or from sphaleron transitions. A difficult problem is how to convert
the estimates for generating non-zero topological charge into one for
$P$ and $CP$ odd asymmetries for measured 
pions~\footnote{We speculate that one may construct, akin to $\vec{L}\cdot 
\vec{S}$ coupling, invariants resulting 
from the coupling of the vorticity of CS number to that of 
the vorticity of collective flow in an expanding plasma.}.  
We will not address
this problem directly here.  Instead, since the quantity $\langle\nu^2\rangle$
that we compute here is closely related to the topological
susceptibility, we can relate our estimate to those of
Refs.~\cite{KPT,KP}.  Our estimates here suggest that the contribution
to $a$ from classical fields is small and would not therefore affect
the estimate of Ref.~\cite{KP}. On the other hand, the contribution
from sphaleron transitions might be potentially large.  Thus if the
RHIC experiments see a large $P$ and $CP$ violating effect, it may also be due to 
sphalerons~\cite{Shuryak}.

Our results depend on the strong boost invariance assumption, namely, each 
event is boost invariant.
At central rapidities, the measured spectra look boost invariant~\cite{PHBR}
 on average. However, what is relevant to our assumption is the 
requirement that the spectra are boost invariant on an event by event basis.
One way to ensure this is to compute the standard deviation from strict
boost invariance in one unit of rapidity about mid-rapidity and check if
it is small. Such 
studies are underway but no conclusive results have been published 
yet~\cite{Baker}. As one goes to larger rapidities, the boost invariance 
assumption breaks down. The no-go theorem forbidding sphaleron transitions is 
violated and one may therefore see larger effects at non-central rapidities 
(in the $P$ and $CP$ 
odd observables currently being investigated) due to sphaleron transitions. 
It will be very interesting to see if this prediction can be tested  
in the on-going experiments at RHIC.

\section*{Acknowledgments}

We thank Dietrich B\"{o}deker, Greg Carter, Yuri Kovchegov,
Eugene Levin, Larry McLerran,
Emil Mottola, Rob Pisarski, Mikhail Shaposhnikov  
and Edward Shuryak for useful discussions. D.K. and
R.V. were supported by DOE Contract No. DE-AC02-98CH10886. A.K. and
R.V acknowledge the support of Portuguese FCT, under grants 
CERN/P/FIS/15196/1999 and CERN/P/FIS/40108/2000. Finally, R.V. would like to
acknowledge the
support of RIKEN-BNL, and A.K. would like to thank the Nuclear Theory Group, 
BNL, for its hospitality.

\end{document}